
\magnification=1200
\baselineskip=14pt
\tolerance=10000

\hfill\break
\rightline{NWU-6/94}
\rightline{September 1994}
\vskip 0.5cm
\line{\hfil {\bf Comment on \lq \lq Monte Carlo
Simulation of a First-Order Transition} \hfil}
\line{\hfil {\bf for Protein Folding"} \hfil}
\vskip 1.5cm

\centerline{Bernd A. Berg,$^1$ Ulrich H.E.~Hansmann,$^2$
and Yuko Okamoto$^3$}
\vskip 0.5cm

\centerline{$^1${\it Department of Physics, Florida State University,
            Tallahassee, FL 32306, USA}}
\vskip 0.5cm

\centerline{$^2${\it Konrad Zuse Zentrum f{\"u}r Informationstechnik
Berlin (ZIB) }}
\centerline{{\it  10711 Berlin, Germany}}
\vskip 0.5cm

\centerline {$^3${\it Department of Physics,
Nara Women's University, Nara 630, Japan}}

\vskip 1.5cm

The article by Hao and Scheraga [1] states that it is the first
application of the so-called \lq \lq entropic sampling Monte Carlo
(ESMC)" method [2] to the protein folding problem:
\lq \lq It should be noted that the earlier procedures are all
aimed at simple systems such as spin lattices and fluids".  However,
there already exists an earlier work that introduced the same method
to the protein folding problem [3]. In the latter article the
method was referred to as \lq \lq multicanonical Monte Carlo (MMC)"
[4-7].  Although MMC was quoted in Ref.~[1], Hao and Scheraga failed
in recognizing the identity of ESMC to the earlier MMC method. The
reason is possibly that the authors of Refs. [1,2] seem
not to have been aware
of Refs. [6,7], where the  idea and numerical methodology
 had been more
explicitly stated than in Refs. [4,5].
These latter works (Refs. [6,7]) had already
appeared in print before Ref.~[2] was submitted for publication. The
erratum of Ref.[2] corrects for this neglect, but
Hao and Scheraga [1] do not refer to this erratum.
It is the purpose of this comment to give a simple proof that
the two superficially differently looking approaches are identical.
Essentially, we follow some earlier private correspondence of one of
the authors with Lee [8].

In ESMC configurations with energy $E$ are updated with a weight
$$ w(E)\propto e^{-S(E)}~,\eqno(1)$$
(see, for example, Eq.~(3) of Ref.~[1]) where $S(E)$ is
the microcanonical
entropy. We have set the Boltzmann constant
$k$ equal to 1 for simplicity.
On the other hand, in MMC a simulation is performed in a
{\it multicanonical ensemble} [5,7], which in turn is {\it defined} by
the condition that
the probability distribution of the energy shall be constant:
$$ P(E) = n(E)w(E) = {\rm const}~, \eqno(2)$$
(see, for example, Eq.~(4) of Ref.~[6]) where $n(E)$ is the
spectral density (or density of states).
Hence, this condition implies a weight factor
$$w(E) \propto n^{-1} (E)~.\eqno(3)$$
It is standard textbook knowledge of statistical mechanics that the
density of states $n(E)$ and the entropy $S(E)$ are related by
$$n(E) = e^{S(E)}~. \eqno(4)$$
Hence, ESMC and MMC have the identical weight factor $w(E)$
(see Eqs. (1), (3), and (4)).
Both methods are based on the same idea that a uniform (flat)
distribution in energy is obtained (rather than the usual bell-shaped
canonical distribution) so that even regions with small
$n(E)$ or $S(E)$ may be explored in detail and thus a good estimate
of the density of states or entropy may be obtained [2,4-7].
Therefore, the two methods are
conceptually identical.

The identity of the two methods was obscured possibly because
most works based on
MMC use the following
parametrization of the weight factor of Eq.~(3):
$$w(E)\propto e^{-\beta(E)E -\alpha(E)},\eqno(5)$$
which is different from that of ESMC (see Eq.~(1)).
Here, $\beta (E)$ can be interpreted as an  ``effective'' inverse temperature.
While not
necessary, this choice is particularly appealing when one wants to
combine canonical and multicanonical simulations [3--7]. At first sight,
one may be puzzled by the fact that ESMC has only one parameter $S(E)$,
while MMC seems to have two parameters $\alpha(E)$ and $\beta(E)$.
However, there is also only one free parameter for MMC,
since $\alpha(E)$
is obtained from $\beta(E)$ by the relation [4,6]
$$\alpha(E)=\alpha(E') + (\beta(E') - \beta(E)) E~,\eqno(6)$$
where $E'$ and $E$ are adjacent bins in the histogram. Both ESMC
and MMC discretize the energy space with a certain bin size and
consider the (unnormalized) histogram $H(E)$ of the energy distribution.
The parameters $S(E)$ for ESMC and $\beta(E)$ for MMC are both
determined by an iteration of simulations with small Monte Carlo statistics.
In ESMC $S(E)$ is updated by [2]
$$S^{i+1} (E) = S^i(E) + \ln H^i (E)~~~ {\rm for}~~ H^i (E) \ge 1, \eqno(7a)$$
and
$$S^{i+1} (E) = S^i(E) ~~~ {\rm for}~~ H^i (E) = 0 . \eqno(7b)$$
Here, $i$ is the iteration number and $H^i (E)$ is the histogram
of the energy distribution after $i$-th iteration.

We now show that the recursion relation, Eq. (7a), for ESMC is
identical with that for MMC
derived earlier by Berg and Celik [6].
The identity of
the weight factors of both methods implies the following relation
among the parameters at the $i$-th iteration (see Eqs. (1) and (5)):
$$S^i(E) = \beta^i(E) E + \alpha^i(E)~. \eqno(8)$$
By Eq. (6) we can eliminate $\alpha^i(E)$ and obtain
$$S^i(E') - S^i(E)=\beta^i(E') (E' - E)~. \eqno(9)$$
This means that once we have ESMC parameter $S(E)$, we can obtain
MMC parameters $\beta(E)$ and $\alpha(E)$, and vice versa.
Indeed, Eq. (9) with the recursion relation Eq. (7a) gives the
following relation
$$ \beta^{i+1} (E') = \beta^i (E') +
{1 \over E' - E}
\ln \left( {H^i(E')\over H^i (E)} \right) , \eqno(10)$$
which is exactly the recursion relation given in
Eq. (5) of Ref.~[6] ($E' - E = -4$ there).
Therefore, the two methods are algebraically and numerically identical.
Step by step
the computer will generate the same configurations, provided that
trivial details like random numbers
 are also chosen to be identical. This is why we continue to
refer to the method as MMC in our subsequent papers [9-11].
\hfil\break

\line{{\bf References and Notes} \hfil}
\item{[1]} Hao, M.-H.; Scheraga, H.A. {\it J. Phys. Chem.} {\bf 1994},
{\it 98}, 4940.

\item{[2]} Lee, J. {\it Phys. Rev. Lett.} {\bf 1993}, {\it 71},
211; Erratum {\bf 1993}, {\it 71}, 2353.

\item{[3]} Hansmann, U.; Okamoto, Y. {\it J. Comput. Chem.} {\bf 1993},
{\it 14}, 1333.

\item{[4]} Berg, B.; Neuhaus, T. {\it Phys. Lett.} {\bf 1991},
{\it B267}, 249.

\item{[5]} Berg, B.; Neuhaus, T. {\it Phys. Rev. Lett.} {\bf 1992},
{\it 68}, 9.

\item{[6]} Berg, B.; Celik, T. {\it Phys. Rev. Lett.} {\bf 1992},
{\it 69}, 2292.

\item{[7]} Berg, B. {\it Int. J. Mod. Phys.} {\bf 1992}, {\it C3},
1083.

\item{[8]} Berg, B. (unpublished, 1993).

\item{[9]} Okamoto, Y.; Hansmann, U. {\it Predicting $\alpha$-helix
propensities of nonpolar amino acids by multicanonical algorithm},
preprint FSU-SCRI-94-28 and NWU-1/94.

\item{[10]} Hansmann, U.; Okamoto, Y. {\it Sampling ground-state
configurations of a peptide by multicanonical annealing},
preprint FSU-SCRI-94-30 and NWU-2/94, {\it J. Phys. Soc. Jpn.}
{\bf 1994}, {\it 63}, in press.

\item{[11]} Hansmann, U.; Okamoto, Y. {\it Comparative study of
multicanonical and simulated annealing algorithms in the protein
folding problem},
preprint SC-94-20 and NWU-5/94, {\it Physica A}, in press.
\hfil\break
\vfill
\bye
\end